\documentclass[aps,prb,reprint,superscriptaddress,amsmath,amssymb]{revtex4-2}
\usepackage[T1]{fontenc}
\usepackage{microtype}
\usepackage{graphicx}
\usepackage{bm}
\usepackage{booktabs}
\usepackage{xcolor}
\usepackage[colorlinks=true,citecolor=blue,linkcolor=blue,urlcolor=blue]{hyperref}
\usepackage{soul}
\usepackage{etoolbox}
\sethlcolor{yellow}
\soulregister{\cite}{7}
\soulregister{\ref}{7}
\soulregister{\eqref}{7}
\robustify{\hl}
\newcommand{\kap}{\kappa}
\newcommand{\teff}[1]{\tau_{\mathrm{eff},#1}}
\newcommand{\unitsk}{\mathrm{W\,m^{-1}\,K^{-1}}}
\begin{document}
\title{Anisotropic pressure response of thermal conductivity in crystalline polytetrafluoroethylene}
\author{Maissa Adi}
\affiliation{Department of Chemical Engineering, McGill University, Montreal, Canada}
\author{Jamal Abou Haibeh}
\affiliation{Department of Chemical Engineering, McGill University, Montreal, Canada}
\author{Samuel Huberman}
\email{samuel.huberman@mcgill.ca}
\affiliation{Department of Chemical Engineering, McGill University, Montreal, Canada}
\affiliation{Department of Physics, McGill University, Montreal, Canada}
\begin{abstract}
Compression can change thermal conductivity differently along polymer chains and across them by altering phonon velocities and lifetimes. We examine this directional response in orthorhombic Form III polytetrafluoroethylene (PTFE) at calculated pressures of 1.503, 3.452, and 5.056 GPa, using first-principles harmonic and cubic force constants within the single-mode relaxation-time approximation (SMRTA). At 100, 200, and 300 K, the chain-axis conductivity decreases at every pressure step. Between 3.452 and 5.056 GPa, both transverse components recover: at 300 K they increase by $23.7\%$ and $13.7\%$, while the longitudinal conductivity decreases by $14.5\%$. An exact factorization separates the conductivity into a heat-capacity--velocity factor and a weighted effective lifetime. The transverse recovery follows from larger heat-capacity-weighted squared velocities, consistent with stiffer interchain restoring forces. Along the chains, shorter effective lifetimes dominate, and both conductivity-weighted Normal and Umklapp rates increase at 300 K. Comparing the same cubic force-constant blocks at all three pressures, we find that interchain coupling is already substantial at 1.503 GPa and increases under compression.
\end{abstract}
\maketitle
\raggedbottom

\section{Introduction}
Crystalline polymers conduct heat through vibrations that sample strong covalent bonds along the chains and weaker interactions between them. For crystalline polyethylene (PE), first-principles conductivities are of order $10^2~\unitsk$ along the chains but only $0.5\text{--}2.5~\unitsk$ across them, and interchain van der Waals coupling lowers the chain-axis value by opening scattering channels \cite{Wang2017,Shulumba2017}. Compression changes the restoring forces that set phonon velocities and the anharmonic interactions and phase space that determine phonon lifetimes. The conductivity reflects the competition between these changes. For cubic crystals, calculations predict decreasing conductivity when lifetimes shorten faster than velocities increase \cite{Lindsay2015}, and nonmonotonic responses when these trends compete \cite{Yuan2018}. Across weak van der Waals contacts in layered and molecular crystals, compression generally increases conductivity through larger group velocities \cite{Meng2019,Ye2024,Legenstein2026}.

High-pressure Form III polytetrafluoroethylene (PTFE) provides a defined crystal structure in which to examine this competition. Diffraction, Raman spectroscopy, and first-principles calculations characterize Form III as planar zigzag chains packed in an orthorhombic two-chain cell that persists to at least 5 GPa \cite{Flack1972,Nakafuku1975,Rabolt1978,Eby1990,Lorenzen2003,Fatti2019}. In the ambient helical phases, interchain forces influence the vibrational dispersion \cite{Piseri1973}, and the conductivity of bulk Form III PTFE increases with pressure \cite{Kieffer1976}. Molecular-dynamics estimates of the chain-axis conductivity of PTFE range from $3.4~\unitsk$ for a conformationally disordered model \cite{Zhang2014} to $114\text{--}226~\unitsk$ for isolated, straightened chains at $223\text{--}373$ K \cite{Liu2021}, while first-principles calculations have treated PE and partially fluorinated polymer crystals \cite{Utimula2019}. A first-principles calculation including finite-temperature anharmonicity and nuclear quantum effects reported a room-temperature conductivity of $48.3~\unitsk$ for one Form III volume \cite{Cheng2022}. First-principles studies of polymer crystals have also emphasized anisotropic phonon velocities, finite-temperature lattice dynamics, and scattering beyond three-phonon processes \cite{Cheng2019,Cheng2022,Shulumba2017,Reicht2026}. The pressure dependence of each conductivity component provides a way to examine how propagation and relaxation compete within a specified crystal structure.

We calculate thermal conductivity along and across the chains in three orthorhombic Form III structures over 100–300 K. Between 3.452 and 5.056 GPa, compression reduces the chain-axis conductivity while increasing both transverse components. An exact factorization into a heat-capacity–velocity factor and an effective lifetime identifies which terms dominate each response; the directional components weight the same mode lifetimes differently. We further resolve the conductivity-weighted contributions of Normal and Umklapp scattering. Frequency-resolved conductivity and matched phonon branches locate the spectral changes, while harmonic and cubic force constants characterize interchain coupling. We interpret the force-constant norms separately from the scattering rates when connecting these microscopic diagnostics to the transport response.

\begin{figure*}[t]
 \includegraphics[width=\textwidth]{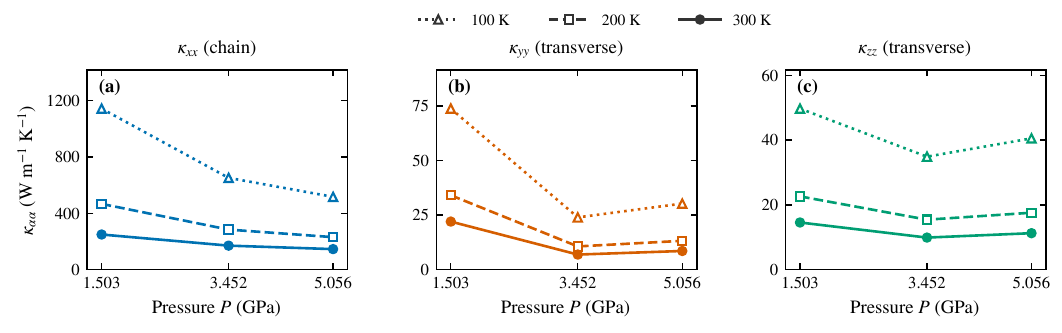}
 \caption{Pressure dependence of the diagonal thermal conductivity of orthorhombic PTFE. (a) Chain-axis component and (b),(c) transverse components at 100 K (open triangles, dotted), 200 K (open squares, dashed), and 300 K (filled circles, solid). Vertical scales differ between panels. Lines connect the calculated states as guides to the eye.}
 \label{fig:tensor}
\end{figure*}

\section{Model and transport analysis}
The orthorhombic cells contain 12 atoms and two chains. We retain the calculation axes, with $x$ along the chains and $y,z$ transverse. In the standard orthorhombic $Pnma$ crystallographic setting, the chain axis is labeled $b$. At 1.503 GPa, $y$ and $z$ are the long ($8.01\,\text{\AA}$) and short ($5.73\,\text{\AA}$) transverse axes. The calculated pressures are 1.503, 3.452, and 5.056 GPa, with cell volumes 120.6965, 115.2100, and 111.9602 \AA$^3$, respectively. Starting from the orthorhombic model of Ref.~\cite{Fatti2019}, we relax the cell shape and atomic positions at fixed volume and identify the pressure from the resulting hydrostatic stress. These volumes agree within about $1\%$ with synchrotron x-ray data for Form III at 294 K and the chain repeat length agrees within $0.4\%$ \cite{Lorenzen2003}. The calculation is restricted to orthorhombic packing; shear-associated monoclinic packing is discussed in Refs.~\cite{Eby1990,Lorenzen2003}.

The DFT calculations use \textsc{Quantum ESPRESSO} \cite{Giannozzi2009}, projector-augmented-wave datasets from pslibrary 1.0.0 \cite{Blochl1994,DalCorso2014}, and the nonlocal optB88-vdW functional \cite{Dion2004,Klimes2010}, selected as \texttt{vdw-df-obk8} and previously used for phonon transport in crystalline PE \cite{Wang2017}. We calculate forces in 504-atom $7\times2\times3$ supercells with electronic $\Gamma$-point sampling. Harmonic force constants are treated with \textsc{phonopy}, and cubic constants are extracted with \texttt{thirdorder.py} \cite{Li2014,Togo2023}. Transport is evaluated with \textsc{phono3py} 3.19.2 \cite{Togo2015,Togo2023}. Pressure-specific electronic settings, phonon meshes, interaction cutoffs, and force-constant sources are documented in the Supplemental Material \cite{Supplement}.

In the single-mode relaxation-time approximation (SMRTA) \cite{Togo2015},
\begin{equation}
 \kap_{\alpha\alpha}=\frac{1}{NV}\sum_\lambda C_\lambda v_{\lambda\alpha}^2\tau_\lambda,
 \label{eq:kappa}
\end{equation}
where $V$ is the primitive-cell volume, $N$ is the number of mesh points, and $\lambda=(\bm q,j)$ identifies a wave vector and branch. The mode heat capacity is $C_\lambda$, the velocity component is $v_{\lambda\alpha}$, and the mode lifetime is $\tau_\lambda$. The sum runs over the full Brillouin zone or, equivalently, over its irreducible part with symmetry weights. Lifetimes include three-phonon and isotope scattering \cite{Tamura1983}. For half linewidths $\gamma$ in ordinary frequency units (THz),
\begin{equation}
 \tau_\lambda^{-1}=4\pi(\gamma_\lambda+\gamma_\lambda^{\mathrm{iso}}),
 \label{eq:rate}
\end{equation}

To compare propagation and relaxation in each direction, we write the conductivity as the product of a heat-capacity--velocity factor $B_\alpha$ and a weighted effective lifetime $\teff{\alpha}$:
\begin{align}
 B_\alpha&=\frac{1}{NV}\sum_\lambda C_\lambda v_{\lambda\alpha}^2
          =C_V\langle v_\alpha^2\rangle_C,\label{eq:B}\\
 \teff{\alpha}&=\frac{\sum_\lambda C_\lambda v_{\lambda\alpha}^2\tau_\lambda}
                         {\sum_\lambda C_\lambda v_{\lambda\alpha}^2},\label{eq:teff}\\
 \kap_{\alpha\alpha}&=B_\alpha\teff{\alpha}.\label{eq:identity}
\end{align}
Here $C_V$ is the volumetric heat capacity, and the subscript $C$ denotes heat-capacity weighting. The factor $B_\alpha$ measures the capacity to carry heat through phonon propagation. The effective lifetime $\teff{\alpha}$ averages the same mode lifetimes in each direction, with weights set by mode heat capacities and squared velocity components. Equation~\eqref{eq:identity} is exact within Eq.~\eqref{eq:kappa}. Pressure changes these weights as well as the mode lifetimes. The factorization therefore quantifies their combined effect on transport; assigning changes uniquely to harmonic or anharmonic interactions requires further analysis. The corresponding conductivity-weighted rate decomposition is
\begin{equation}
 \displaystyle
 \teff{\alpha}^{-1}\simeq\langle\Gamma^{\mathrm{N}}\rangle_\alpha+\langle\Gamma^{\mathrm{U}}\rangle_\alpha
 +\langle\Gamma^{\mathrm{iso}}\rangle_\alpha,
 \label{eq:channels}
\end{equation}
where $\Gamma^{\mathrm{N,U}}_\lambda=4\pi\gamma^{\mathrm{N,U}}_\lambda$ are the Normal and Umklapp three-phonon rates as partitioned by \textsc{phono3py}, $\Gamma^{\mathrm{iso}}_\lambda$ is the isotope rate, and $\langle\cdot\rangle_\alpha$ is an average weighted by the modal contributions to $\kap_{\alpha\alpha}$. The Supplemental Material \cite{Supplement} provides structural comparisons with experiment, numerical tables, frequency-resolved conductivity changes (Fig.~S1), interaction-range data, mesh-refinement checks, mode-map limitations, and consistency checks.

\section{Directional competition and spectral redistribution}
The chain-axis conductivity $\kap_{xx}$ decreases at every sampled pressure step and temperature [Fig.~\ref{fig:tensor}]. At 300 K it falls from 249.72 to 170.88 and then 146.15 $\unitsk$. Both transverse components instead decrease between 1.503 and 3.452 GPa and increase between 3.452 and 5.056 GPa. Their 300 K sequences are $22.05\rightarrow6.94\rightarrow8.59$ $\unitsk$ for $y$ and $14.58\rightarrow9.92\rightarrow11.28$ $\unitsk$ for $z$. The transverse recovery occurs at all three temperatures, with the high-pressure values remaining below the low-pressure values. We focus the transport-balance interpretation on the recovery between 3.452 and 5.056 GPa.

\begin{figure*}[!t]
 \centering
 \includegraphics[width=\textwidth]{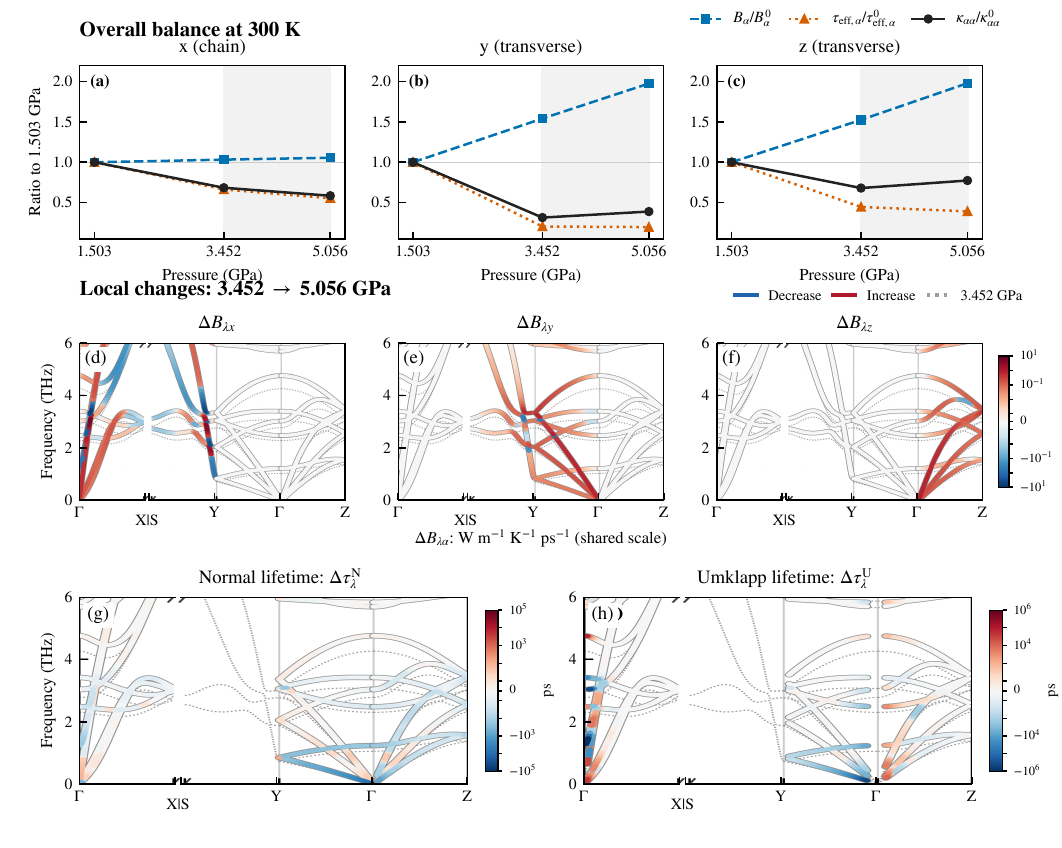}
 \caption{Overall transport balance and local mode changes at 300 K. (a)$\text{--}$(c) Conductivity, heat-capacity$\text{--}$velocity factor $B_\alpha$, and effective lifetime along $x$, $y$, and $z$, each normalized to its value at 1.503 GPa (superscript 0); their product relation is Eq.~(5). Shading marks the interval used for the mode maps. (d)$\text{--}$(f) Changes in mode factors $B_{\lambda\alpha}=C_\lambda v_{\lambda\alpha}^2/V$ for $x$, $y$, and $z$, on a shared color scale. (g),(h) Changes in Normal and Umklapp lifetimes $\tau^{\mathrm{N,U}}_\lambda=(4\pi\gamma^{\mathrm{N,U}}_\lambda)^{-1}$, with separate scales in ps. Mode maps show 5.056 GPa minus matched 3.452 GPa values along $\Gamma\text{--}\mathrm{X}|\mathrm{S}\text{--}\mathrm{Y}\text{--}\Gamma\text{--}\mathrm{Z}$ on the 5.056 GPa dispersion; dotted gray curves show 3.452 GPa. Blue and red denote decreases and increases, on symmetric logarithmic scales. Panels (a)$\text{--}$(c) use full-zone data; the mode maps illustrate local branch changes. Averaging mode factors over the full-zone mesh gives $B_\alpha$. The channel lifetimes in (g),(h) differ from the directional effective lifetimes in (a)$\text{--}$(c). Assignment, lifetime interpolation, and color coverage are detailed in the Supplemental Material, where Fig.~S3 retains all pressure comparisons.}
 \label{fig:modes}
\end{figure*}

At 300 K, shorter effective lifetimes dominate the chain-axis response, while growth of the transverse transport factors drives the higher-pressure recovery [Fig.~\ref{fig:modes}(a)$\text{--}$(c)]. Along $x$, $B_x$ increases by only 5.57\% over the complete interval, whereas $\teff{x}$ decreases by 44.56\%. The small increase in $B_x$ follows from densification: the inverse cell volume increases by $7.80\%$, whereas the heat capacity per cell and $\langle v_x^2\rangle_C$ decrease by $1.3\%$ and $0.8\%$; the latter corresponds to a change in the heat-capacity-weighted root-mean-square chain velocity from $2.731$ to $2.720~\mathrm{km\,s^{-1}}$. Thus, the all-frequency chain-axis loss is dominated by shorter effective lifetimes rather than a uniform decrease in velocity.

The transverse factors $B_y$ and $B_z$ increase much more strongly than $B_x$. During the first pressure step they rise by 54.05\% and 52.39\%, but the corresponding effective lifetimes fall by 79.56\% and 55.35\%. Within the calculated model, relaxation therefore dominates the initial transverse suppression. During the second pressure step, propagation gains outweigh relaxation losses: $B_y$ and $B_z$ rise by 28.14\% and 29.79\%, while the effective lifetimes decrease by only 3.46\% and 12.40\%. Their products increase by 23.70\% and 13.70\%. In the same interval, $B_x$ rises by 2.30\% and $\teff{x}$ falls by 16.40\%, giving a further 14.47\% reduction of $\kap_{xx}$.

Larger transverse squared velocities drive most of the higher-pressure increase in $B_\alpha$. The volumetric heat capacity increases by 2.32\%, whereas the heat-capacity-weighted squared velocities $\langle v_y^2\rangle_C$ and $\langle v_z^2\rangle_C$ increase by 25.23\% and 26.85\%. The increase in squared velocities therefore outweighs the decrease in all-frequency effective lifetimes. At 100 K, $\teff{y}$ instead increases slightly over this interval; the common feature across temperatures is the substantial growth of transverse transport factors. A comparable velocity-driven increase occurs in the cross-plane conductivity of compressed layered and molecular crystals: in MoS$_2$ it increases from 3.5 to about $25~\unitsk$ although the longitudinal-acoustic lifetime decreases about threefold \cite{Meng2019}, whereas in graphite and naphthalene the weak-direction lifetimes also increase \cite{Ye2024,Legenstein2026}. PTFE, with transverse effective lifetimes changing by only $-3.5\%$ and $-12.4\%$, lies between these limits.

Resolving the conductivity by frequency distinguishes changes within a spectral sector from redistribution between sectors. Across the full pressure range, the chain-axis conductivity below 3 THz decreases by 112.38 $\unitsk$, exceeding the net loss of 103.57 $\unitsk$ because higher-frequency contributions increase. Within this low-frequency sector, $B_x$ decreases from 2.019 to 1.536 $\unitsk\,\mathrm{ps}^{-1}$ and the effective lifetime decreases from 104.61 to 64.33 ps. Both factors suppress the low-frequency chain contribution. The increase in the all-frequency $B_x$ therefore includes redistribution toward higher-frequency modes. Modes below 3 THz carry $84.6\%$, $74.3\%$, and $67.6\%$ of $\kap_{xx}$ at the three pressures, and $77\%\text{--}85\%$ of $\kap_{xx}$ comes from modes with $|v_x|>5~\mathrm{km\,s^{-1}}$. In PE, chain-axis heat transport is dominated by $11\text{--}16$ THz longitudinal phonons whose main three-phonon decay channel is closed \cite{Reicht2026}. In PTFE, chain-axis heat is carried mainly by low-frequency, acoustic-like phonons that overlap the rigid-chain optical manifold.

The transverse recovery is concentrated at low frequencies: $95.6\%$ of the $y$ increase and $93.7\%$ of the $z$ increase lie below 5 THz [Fig.~S1(b),(c)]. The 2--5 THz region supplies approximately 69\% and 72\% of these gains. In the 3--5 THz interval, $B_y$ and $B_z$ rise by about 65\% and 78\%, accompanied by modest effective-lifetime increases.

Figure~\ref{fig:modes} connects the overall balance in panels (a)$\text{--}$(c) to local mode changes during the higher-pressure recovery in panels (d)$\text{--}$(h). We match branches between pressures by eigenvector overlap; Fig.~S3 in the Supplemental Material gives the maps for all three pressure comparisons \cite{Supplement}. At $\Gamma$, the eight lowest modes are rigid-chain translations and rotations; the relative translations along $y$ and $z$ harden from 3.49 to 4.75 THz and from 2.38 to 3.03 THz over the pressure series, consistent with $K_y$ and $K_z$ (Sec.~\ref{sec:fc}). In the higher-pressure interval, the maps show increases in $B_y$ and $B_z$ on many branches below 5 THz, with lifetime changes of both signs. Along $\Gamma\text{--}\mathrm{X}$, $B_x$ changes little and with both signs.

\begin{figure*}[!t]
 \includegraphics[width=\textwidth]{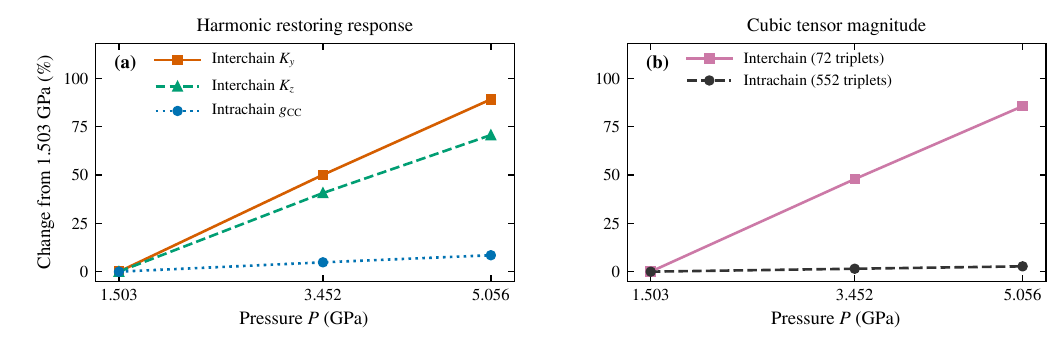}
 \caption{Pressure dependence of real-space force-constant diagnostics. (a) Relative-chain translation curvatures $K_y,K_z$ [Eq.~\eqref{eq:K}] and the nearest-neighbor C--C bond projection $g_{\mathrm{CC}}$. Each probe is normalized to its own low-pressure value. (b) Changes in the cubic norm $F_3$ [Eq.~\eqref{eq:fc3}] for identical triplet sets at all pressures. Panel (b) compares 552 intrachain and 72 interchain blocks common to all three pressures. Purely onsite terms are excluded. These tensor norms quantify real-space cubic coefficients.}
 \label{fig:fc}
\end{figure*}

\section{Harmonic and cubic coupling between chains}
\label{sec:fc}
The transverse velocity gains are consistent with stronger interchain restoring forces. We examine this connection with a harmonic probe: we displace the two chain sublattices rigidly by $+\delta/2$ and $-\delta/2$ along a transverse axis and evaluate the $\Gamma$-point energy curvature per primitive cell,
\begin{equation}
 K_\alpha=\bm U_\alpha^{\mathsf T}\bm\Phi^{(2)}_\Gamma\bm U_\alpha,
 \qquad \alpha=y,z.
 \label{eq:K}
\end{equation}
Here $\bm U_\alpha$ contains the signed half-unit displacements. Across the pressure series, $K_y$ increases by 89.1\% and $K_z$ by 70.7\% [Fig.~\ref{fig:fc}(a)]. These rigid-sublattice probes evaluate restoring curvatures with fixed internal coordinates. The nearest-neighbor C--C projection $g_{\mathrm{CC}}=-\bm n^{\mathsf T}\bm\Phi^{(2)}_{ij}\bm n$, with $\bm n$ along the bond, increases by 8.50\%. Compression thus strengthens the relative-chain restoring response and this measure of backbone coupling.

To compare cubic coupling at fixed triplet membership, we evaluate the tensors over a set $S$ of ordered atomic triplets using
\begin{equation}
 F_3(S)=\left[\frac{1}{N_S}\sum_{(ijk)\in S}
              \sum_{abc}|\Phi^{(3)}_{ia,jb,kc}|^2\right]^{1/2}.
 \label{eq:fc3}
\end{equation}
This root-mean-square tensor norm avoids cancellation of signed Cartesian components. Triplets are classified by physical chain identity, including their lattice images. The common set contains 552 offsite intrachain and 72 interchain blocks; purely onsite blocks are excluded. The interchain $F_3$ increases from 1.035 to 1.921 eV \AA$^{-3}$, or $85.6\%$, while the intrachain value increases from 61.584 to 63.278 eV \AA$^{-3}$, or $2.75\%$ [Fig.~\ref{fig:fc}(b)]. All common interchain blocks involve fluorine displacements on two chains. A subset of 24 interchain $\mathrm{F}\text{--}\mathrm{F}\text{--}\mathrm{F}$ blocks has an RMS norm of 1.454, 2.106, and 2.607 eV \AA$^{-3}$ at increasing pressure. These interactions are therefore already substantial at low pressure and strengthen under compression.

The transverse recovery coexists with growth of the interchain cubic norm, so a decrease in this norm cannot explain it. Scattering depends on three-phonon amplitudes, obtained by projecting the cubic tensors onto phonon eigenvectors and summing coherently with mass and frequency factors. Scattering rates also depend on occupation and conservation constraints \cite{Togo2015}. Intrachain and interchain amplitudes can interfere, and either can scatter a mode whose heat current is primarily longitudinal. The directional contrast thus reflects different transport balances, with modes in each direction sampling both intra- and interchain interactions.

Equation~\eqref{eq:channels} partitions the effective relaxation rate. Along $x$ at 300 K, $\langle\Gamma^{\mathrm{N}}\rangle_x$ increases from $0.033$ to $0.049$ and $0.058~\mathrm{ps^{-1}}$ and $\langle\Gamma^{\mathrm{U}}\rangle_x$ from $0.040$ to $0.060$ and $0.073~\mathrm{ps^{-1}}$, with isotope scattering below $1\%$. Both conductivity-weighted averages grow in each interval. Because the weights change with pressure, this trend characterizes the averages rather than every individual mode. At 300 K, Normal processes supply $44\%\text{--}61\%$ of the summed conductivity-weighted three-phonon rates across the three directions and pressures. These averages show comparable Normal and Umklapp contributions. Assessing phonon hydrodynamics requires mode-resolved and collision-operator diagnostics \cite{Zhang2020}. Stronger interchain coupling could contribute to the chain-axis loss: in PE, interchain van der Waals coupling lowers the chain-axis conductivity about sixfold relative to an isolated chain by enlarging the three-phonon phase space \cite{Wang2017}, and in multilayer graphene and graphite, interlayer coupling suppresses in-plane transport mainly by relaxing a three-phonon selection rule \cite{Lindsay2011}. The concurrent growth of interchain cubic terms and both weighted scattering-channel averages along $x$ is consistent with this mechanism. Quantifying its contribution requires further mode-resolved analysis.

\section{Discussion and conclusions}
\label{sec:scope}
The pressure response of orthorhombic Form III PTFE reflects a directional competition between phonon propagation and relaxation. Within the static-force-constant SMRTA model, the chain-axis conductivity decreases at every sampled pressure step at 100, 200, and 300 K. Both transverse components recover between 3.452 and 5.056 GPa while the chain-axis component continues to decrease. At 300 K, the longitudinal decrease is $14.5\%$, compared with transverse increases of $23.7\%$ and $13.7\%$.

The identity $\kap_{\alpha\alpha}=B_\alpha\teff{\alpha}$ identifies the factors responsible for this contrast. At 300 K in the higher-pressure interval, $B_y$ and $B_z$ increase by $28.14\%$ and $29.79\%$, outweighing effective-lifetime reductions of $3.46\%$ and $12.40\%$. Along the chains, the lifetime reduction instead dominates the small increase in $B_x$. The transverse gains arise mainly from larger heat-capacity-weighted squared velocities, consistent with the increasing relative-chain harmonic stiffness.

Comparing the same cubic force-constant blocks at all three pressures, we find that interchain coupling is already substantial at 1.503 GPa and grows under compression. The transverse recovery therefore coexists with increasing interchain cubic coupling. Along $x$, both conductivity-weighted Normal and Umklapp rates increase at 300 K. The concurrent growth of these rates and the interchain cubic terms is consistent with an interchain contribution to the longitudinal loss. Mode-resolved scattering analysis is needed to quantify that contribution. Computational parameters, dataset provenance, and numerical validation requirements are detailed in the Supplemental Material \cite{Supplement}.

The PTFE results illustrate a broader feature of polymer crystals: interactions between chains affect heat transport both across the chains and along the covalent backbone. For PE, interchain van der Waals coupling enlarges the three-phonon scattering phase space and lowers the chain-axis conductivity relative to an isolated chain \cite{Wang2017}. Comparisons across polymer chemistries likewise show that molecular structure shapes the distribution and lifetimes of heat-carrying modes \cite{Utimula2019,Reicht2026}. The present PTFE calculations illustrate how pressure shifts this balance within one ordered crystal structure. Interpreting thermal anisotropy in polymer crystals therefore requires considering how chain packing affects both phonon propagation and relaxation.

\section*{Data availability}

Public archiving of the underlying force-constant and transport files remains pending.

\begin{acknowledgments}
The authors acknowledge useful discussions with Laurenz Kremeyer. S.H. acknowledges support from the NSERC Discovery Grants Program under Grant No. RGPIN-2021-02957 and FRQNT Nouveau Chercheur No. 341503. This research was enabled in part by computational resources provided by the Digital Research Alliance of Canada. AI tools were used to assist with analysis-code preparation, figure preparation, and manuscript drafting.
\end{acknowledgments}
\bibliography{references}
\end{document}